\documentclass[aps, prd, twocolumn, lengthcheck, superscriptaddress, showpacs, letterpaper, nofootinbib]{revtex4-1}

\usepackage{epsfig}
\usepackage[usenames]{color}
\usepackage{graphicx}
\usepackage{amsmath}
\usepackage{epstopdf}

\def\bi{\bibitem}

\def\la{\langle}\def\ra{\rangle}
\def\be{\begin{eqnarray}}\def\ee{\end{eqnarray}}
\def\lsim{\mathrel{\rlap{\lower3pt\hbox{\hskip1pt$\sim$}}
     \raise1pt\hbox{$<$}}} 
\def\gsim{\mathrel{\rlap{\lower3pt\hbox{\hskip1pt$\sim$}}
     \raise1pt\hbox{$>$}}} 

\allowdisplaybreaks

\begin{document}

\title{Why {\em Explicit} Strangeness Is Not Relevant In Compact Stars}



\author{Mannque Rho}
\affiliation{Institut de Physique Th\'eorique,
	CEA Saclay, 91191 Gif-sur-Yvette c\'edex, France }
\date{\today}

\begin{abstract}

Drawing largely from my work with co--workers, I present arguments that strangeness does not play an {\it explicit} role in compact-star matter. They are based on the skyrmion description of dense matter combined with Wilsoninan renormalization group approach to kaon nuclear interactions. The key idea is that quark degrees of freedom, carrying strangeness, can be traded in by topology for hadron degrees of freedom.

\end{abstract}

\maketitle
\section{Introduction}\label{sec:Intro}

In a recent study of the structure of massive compact stars~\cite{bsHLS}, there was found to be no need to introduce strangeness degrees of freedom, either hadronic or quarkish, in satisfactorily describing the EoS of compact-star  matter at high density. The framework resorted to an effective field theory approach anchored on certain symmetries ``hidden" in QCD, namely, hidden local symmetry  associated with the vector mesons $\rho$ and $\omega$ and hidden scale symmetry associated with the scalar dilaton $\sigma$. No symmetry associated with the $s$ quark was taken into account. The question naturally arises as to whether it is justified to simply ignore strangeness degrees of freedom such as kaon condensation and/or hyperon presence at the density relevant to the interior of massive compact stars in confronting the recently observed  $\sim 2$-solar mass stars~\cite{1.97S,antoniadis13}. 

In standard chiral perturbation approaches in flavor $SU(3)$ as first considered by Kapan and Nelson~\cite{kaplan-nelson} and studied extensively as recounted in numerous reviews, e.g., \cite{BLR}, the $SU(3)$ chiral Lagrangian  would predict, at the mean field level, s-wave anti-kaons (referred to simply as kaons from here on) condensing at a relatively low density, typically at  $\sim 3n_0$. Such a low-density kaon condensation would have a large impact on a variety of astrophysical phenomena including possible low-mass black holes~\cite{BLR}.  The recent discovery of $\sim 2$ solar-mass compact stars~\cite{1.97S,antoniadis13} \index{massive stars! 2-solar mass}  raised the poignant issue as to whether such kaon condensation -- and equally hyperon presence, triggering softness in the EoS as naively suspected,  are not unequivocally ruled out by the observed massive compact stars. Indeed, the verdict pronounced in \cite{1.97S} was that one of the manifestation of strangeness, namely, the kaon condensation, is  ruled out by the observation.

So the question raised is: Is the discovery of  $\sim 2$-solar-mass stars the smoking-gun signal for the death of kaon condensation or equivalently of hyperons in compact star matter?

There are a variety of scenarios in the literature, either invoking or not -- up-to-date unestablished -- repulsive interactions in strange-flavor channels, that claim that  the so-called ``strangeness problem," both hyperon problem and kaon condensation problem,  can be avoided. It is fair however to say that up to date, given the paucity of experimental information and consistent theoretical tools, there are no generally acceptable scenarios.  Thus this issue remains a totally open problem to be resolved.

I would claim that the verdict pronounced in \cite{1.97S}, though it cannot be ruled out, is however too premature at present given the large uncertainty in the theoretical treatment of strangeness in dense matter. The theoretical results the conclusion made in \cite{1.97S} is drawn from is based on naively attaching tree-order chiral Lagrangian treatments of kaon-nucleon interactions to  correlated nuclear dynamics, both  treated separately at a vastly different level of sophistication and with no feedbacks between the two, which could play a crucial role.  Given the preeminent role of the scalar-  and vector-meson degrees of freedom in the EoS for matter at density $n \gsim 2n_0$ as described in \cite{bsHLS}, which is not taken into account in the conventional treatment of kaon condensation, such a hasty verdict should not be taken seriously. As I will argue below, the situation can be a lot subtler than one would summarily think. In fact, there is an indication known since a long time that a unified treatment of {\it both meson and baryon sectors treated on the same footing} is necessary for a reliable description. For example, when strongly repulsive short-range correlations among the nucleons are suitably taken into account, the naive chiral perturbative estimate can be qualitatively wrong~\cite{pandharipande-thorsson}.  Indeed the repulsion operative in the nuclear sector, well recognized in nuclear dynamics, is of the same origin as the attraction between kaons and nucleons responsible for boson condensation and is found to push the condensation threshold density way beyond the density relevant for compact stars. This indicates  both the baryon sector and meson sector and their interconnections, i.e., back-reactions,  must be treated consistently. To the best of my knowledge, such a consistent treatment has not been done up to date.

Within the formalism adopted in \cite{bsHLS}, it should be in principle feasible to approach the problem with the same $V_{lowk}$RG  employed for two flavors,  generalized to three flavors, say, $V_{lowk}^{SU(3)}$RG. It should rely on $V_{lowk}$'s in coupled channels involving hyperons and kaons. The procedure, although complicated and perhaps even daunting, should however be straightforward.  The proper treatment will require $V_{lowk}$'s in the octet baryon sector, octet pseudo-scalar sector and their couplings, suitably taking into account what's known as  intrinsic density dependence (IDD) in the effective Lagrangian inherited from QCD~\cite{bsHLS}. Working this out to confront nature  is however beyond the present capability, mainly due to  the paucity of experimental information that does not allow the large number of ``bare" parameters (with IDDs endowed properly) of three-flavor scale-invariant HLS Lagrangian to be adequately controlled. In what follows, I will work with a drastically simplified version of the approach and try to arrive at a qualitative confirmation that strangeness via hyperons and kaons does not figure in compact-star equation of state. For this I will first argue that kaon condensation and hyperon presence, treated on the same footing,   appear at the same density in the large $N_c$ limit

\section{Kaon condensation vs. hyperons}\label{kcon vs hyperon}
%
So far in the literature, the appearance of hyperons and the onset of kaon condensation are treated separately. With our Lagrangian implemented with both hidden gauge symmetry and scale invariance that I will refer to as $s$HLS with baryons either explicitly included or generated as skyrmions, they  can be and must be treated simultaneously. The observation mentioned above of the effect of short-range correlation in the baryon sector on kaon condensation~\cite{pandharipande-thorsson} illustrates the import of this correlation.

The first problem to resolve is  whether kaons condense before or after the appearance of  hyperons in dense compact-star matter.
At present as far as I know, the only way this problem can be addressed in a tractable approximation in consistency with the basic premise of QCD -- such as large $N_c$ -- is the skyrmion description in which both kaons and hyperons are treated with the same Lagrangian.  This matter can be addressed~\cite{LR-hk} in terms of the successful Callan-Klebanov bound-state model~\cite{callan-klebanov}. In this model, anti-kaons $K^{-}$ are bound to the $SU(2)$ skyrmion to  yield hyperons.  It turns out that this model can interpolate kaons between the chiral limit ($m_K\rightarrow 0$) and the Isgur-Wise heavy-quark limit ($m_K\rightarrow \infty$)~\cite{klebanov-cargese-lecture}, an advantage in handling the kaon mass as it varies with the vacuum change. The model can be applied to nuclear matter by putting  skyrmions on crystal lattice~\cite{Klebanov-crystal}. It was shown~\cite{LR-hk} that put on a crystal lattice,  the energy difference between the lowest-lying hyperon $\Lambda$ and the nucleon $N$ in medium comes out as
 \be
E^\ast_\Lambda -E^\ast_N=\omega_K^\ast + {\cal O}(N_c^{-1})\label{inmedium}
\ee
where the asterisk represents medium-dependence. Note that the in-medium kaon mass is of $O(N_c^0)$ in the $N_c$ counting. It is fortunate that the leading ${\cal O} (N_c)$ term and the flavor singlet ${\cal O} (N_c^0)$ Casimir energy term -- which is extremely difficult to calculate -- cancel out in the difference.

{Now, the $\Lambda$s will appear in the compact star matter when
\be
\mu_e\geq E_\Lambda^\ast -E_N^\ast
\ee
where $\mu_e$ is the electron chemical potential which is equal to $\mu_n-\mu_p$  in weak equilibrium}. On the other hand, kaons will appear by the weak process $e^-\rightarrow K^- +\nu_e$ when
\be
\mu_e\geq \omega_K^*=m_K^\ast
\ee
for the s-wave kaon. Therefore to the leading order in $N_c$ in QCD, hyperons and condensed kaons populate compact stars simultaneously. Which one appears first in the single Lagrangian description then depends on ${\cal O} (1/N_c)$ hyperfine corrections, namely, when the skyrmion-kaon system is rotationally quantized. A simple quasi-particle approximation leads to
\be
E_\Lambda^\ast-E_N^\ast=\omega^\ast_K +\frac{3}{8\Omega^\ast}({c^\ast}^2-1)\label{med-hyper}
\ee
where $\Omega>0$ is the moment of inertia of skyrmion rotator and $c^*$ is the in-medium hyperfine coefficient multiplying the effective spin operator of strangeness -1. The coefficient $c$ is highly model-dependent even in the matter-free space~\cite{callan-klebanov}, so it is  unknown in dense matter except in the large $N_c$ limit and also in the chiral limit. In either or both of these limits,  $c^\ast\rightarrow 1$. In the matter-free space, it is found to be $c^2\sim 0.5$. Although presently there is no proof, it seems likely that ${c^\ast}^2 <1$ in medium, approaching 1 from below near chiral restoration. If this is the case, that would suggest that hyperons appear before kaons condense
and they ultimately join in the vicinity of chiral restoration. It is however difficult to be more precise on this point since the effect is at ${\cal O} (N_c^{-1})$ and at that order there are many other corrections, such as higher-order nuclear correlations, that go beyond the mean-field order, something that can be done in the three-flavor $V_{lowk}$ approach. In the absence of a realistic calculation with kaon mass and $1/N_c$ corrections, it seems reasonable to  assume that kaon condensation and hyperons appear at about the same density.
\section{Hyperon problem}\index{hyperon problem}
It will be shown below that kaon condensation will {\it inevitably} take place at some high density above $n_0$. The precise value for the density at which kaons start condensing cannot be pinned down theoretically unless a full 3-flavor $V_{lowk}$RG analysis is performed. Here I develop an argument based on a mean-field treatment of $bs$HLS that hyperons may not destabilize massive stars.  With the possibility that kaons condense at about the same density as discussed above, the same will apply to this process. 

It is easy to see by simple energetic considerations that hyperons could be present at high density in compact-star matter. Specifically, the lowest-lying hyperon $\Lambda$, with its attractive interaction, is estimated to appear at matter density $\sim 2n_0$ with the others possibly appearing at higher density. This suggests  that the hyperons could appear at about the same density as the one at which the half-skyrmion phase appears in the skyrmion matter~\cite{bsHLS}. If this were the case, then the prediction made in \cite{bsHLS}  would make no sense without the hyperonic degree of freedom duly taken into account.

In an admittedly phenomenological, but sophisticated, study using a Monte Carlo simulation over parameters that enter in the EoS for symmetric and asymmetric nuclear matter such as the compression moduli $K$ and $L$, symmetry energies $E_{sym}$ and $E_{sym}^\Lambda=S_\Lambda$, Bedaque and Steiner obtain the range of density $\Delta$ constrained by hydrodynamic stability of the system that ensures that stars with $M >2M_\odot$ could be supported~\cite{bedaque-steiner}.  The $\Delta$ is then the range of density within  which the in-medium $\Lambda$ mass should become greater than the vacuum value. One expects -- and it is confirmed experimentally -- that the $\Lambda$-nucleon interaction is attractive at normal nuclear matter density, so $\Lambda$s could  be bound in nuclear matter. In compact star matter, as density increases, the chemical potential difference between neutron and proton increases, and it can become energetically favored to have spontaneous creating of hyperons in the system. This can happen when density reaches roughly twice the normal nuclear matter density. The instability that could be generated by the presence of hyperons at  such low density is  the ``hyperon problem." The $\Delta$ then stands for the range of density at which the $\Lambda$-interactions must be repulsive enough to make the in-medium $\Lambda$ mass be greater than the vacuum mass. The analysis by Bedaque and Steiner establishes that  the range of $\Delta$ required is $1<  \Delta/n_0\lsim3$. This range will be referred to as ``Bedaque-Steiner constraint."

One can  address this problem using the single-decimation RG procedure~\cite{BR:DD} as an approximation to $V_{lowk}^{SU(3)}$RG. One can take the mean-field approximation with the three-flavor baryonic $bs$HLS endowed with the IDDs -- assumed to be $SU(3)$ symmetric for a qualitative treatment -- as equivalent to Fermi-liquid fixed point theory. This approach has been shown to be surprisingly successful for two-flavor systems~\cite{CND-III}, e.g., the anomalous gyromagnetic ratio in heavy nuclei $\delta g_l$~\cite{FR96}, the ``quenched $g_A$ problem~\cite{gA} etc., so it is hopeful that it can give a trustful result in this three-flavor problem.

\begin{figure}[!t]
\begin{center}
\includegraphics[width=0.25\textwidth]{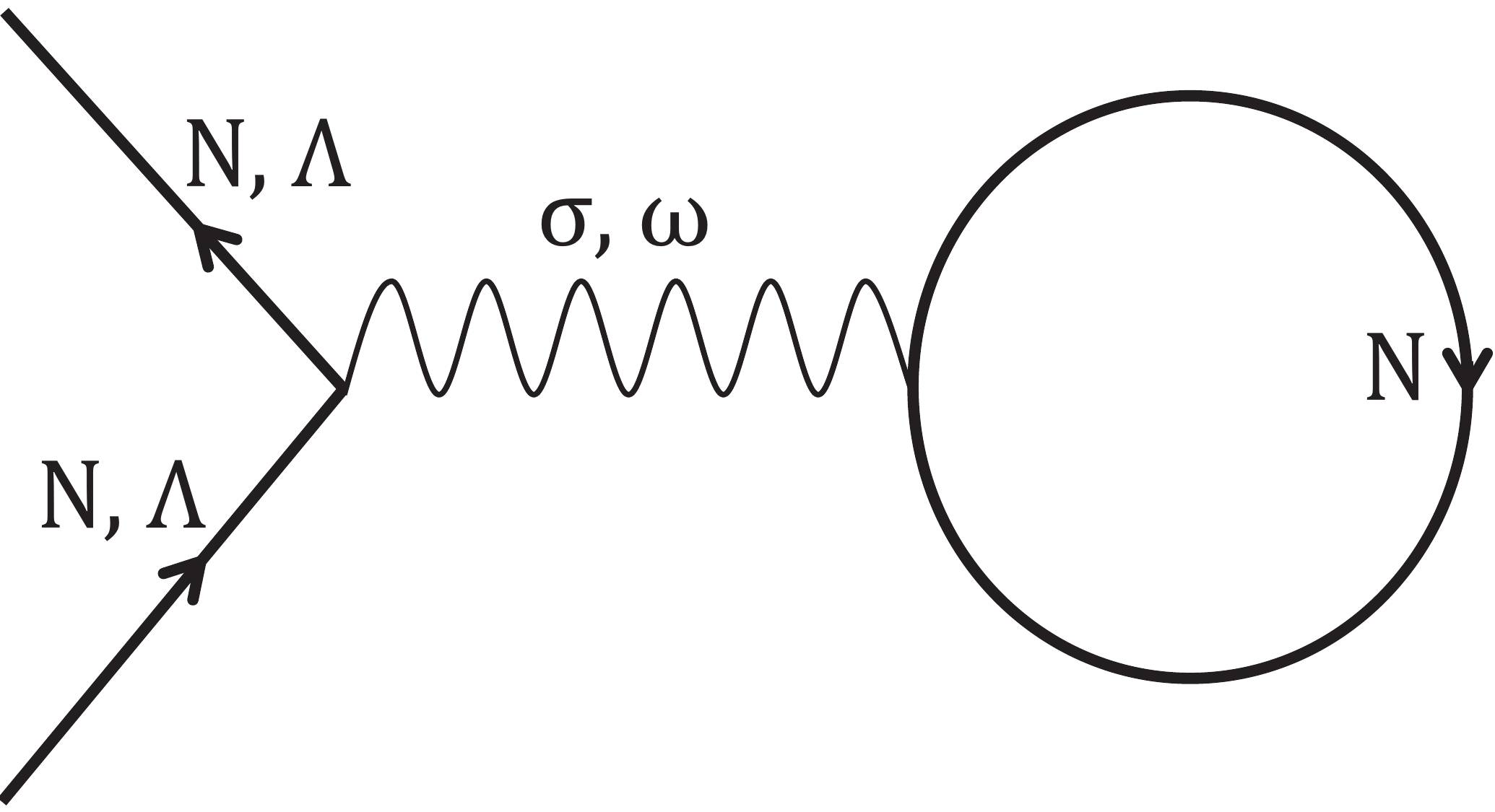}
\caption{Tadpole diagram for self-energies for the nucleon $N$ and the hyperon $\Lambda$ in medium with the IDDs. The loop corresponds to the nucleon scalar density $n_s=\la\bar{N}N\ra$ for coupling to $\sigma$ and the nucleon number density $n=\la N^\dagger N\ra$ for coupling to $\omega$.}
\label{tadpole}
\end{center}
\end{figure}

In the mean field, the chemical potential for the $\Lambda$ in medium gets contributions from two sources, one from the IDD in the ``bare'' mass parameter $m^*_\Lambda\approx \Phi m_\Lambda$ and the other from the potential terms coming from $\Lambda$-nuclear coupling via $\sigma$ and $\omega$ exchanges as depicted in Fig.~\ref{tadpole},
\begin{eqnarray}
\mu_\Lambda
= m^*_\Lambda -\frac{g^\ast_{\sigma\Lambda}g^\ast_{\sigma N}}{m^{\ast\,2}_{\sigma}} n_s +\frac{g^\ast_{\omega \Lambda}g^\ast_{\omega N}}{m^{\ast\,2}_{\omega}} n\, \label{muL2}
\end{eqnarray}
where $n_s$ and $n$ are, respectively, nucleon scalar density and nucleon number density as defined in Fig.~\ref{tadpole}. The notations for $(\sigma,\omega)$ coupling to $\Lambda$ and $N$ are self-evident. The asterisk stands for IDD-scaling parameters.

\begin{figure}[ht]
\begin{center}
\includegraphics[width=8.0cm]{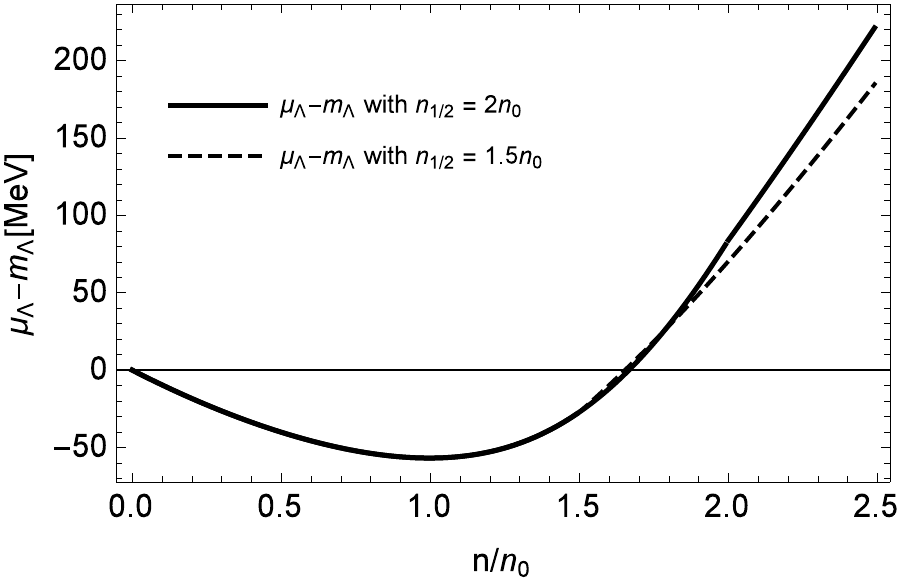}
\caption{$\mu_\Lambda -m_\Lambda$ vs. $n/n_0$ calculated with the scaling parameters determined in the theory and summarized in \cite{bsHLS}. The demarcation density was chosen for $n_{1/2}= (1.5, 2.0)n_0$. The density at which the mass shift crosses zero corresponds to $\Delta$ of \cite{bedaque-steiner}.}
\label{mu_L_PKLR}
\end{center}
\end{figure}
Apart from the $\Lambda$ coupling to the mesons, the large cancelation between the dilaton ($\sigma$) attraction and the vector ($\omega$) repulsion responsible for small binding energy for nuclear matter must take place also in this case. In fact, using the standard constituent quark (equivalent to quasi-quark) counting,  we may take $g_{\sigma \Lambda}\approx \frac 23 g_{\sigma N}$ and $g_{\omega \Lambda}\approx \frac 23 g_{\omega N}$, the 2/3 factor accounting for the two non-strange quarks in $\Lambda$ vs. 3 in nucleon. Then
\begin{eqnarray}
\mu_\Lambda
= m^*_\Lambda +\frac 23 \left(-\frac{g^{\ast\,2}_{\sigma N}}{m^{\ast\,2}_{\sigma}} n_s + \frac{g^{\ast\,2}_{\omega N}}{m^{\ast\,2}_{\omega}} n\right)\, . \label{muL3}
\end{eqnarray}

To make a numerical estimate of the $\Lambda$ mass shift in dense medium, we take into account the IDD-scaling -- assuming flavor $SU(3)$ symmetry -- of the parameters in the $s$HLS Lagrangian with baryons incorporated explicitly (call it $bs$HLS) in the mean-field calculation which corresponds to  the ``single-decimation procedure" of \cite{BR:DD}. In doing this, one should recognize that the scaling parameter  $c_I$ in this procedure could well be different, i.e., renormalized, from the IDD coefficient entering into the double-decimation procedure with $V_{lowk}$ employed in \cite{bsHLS} addressed to compact stars. This is because  in the single-decimation procedure, the scaling function $\Phi_I$ is related to the Fermi-liquid fixed-point parameters as shown in \cite{FR96} and encodes certain nonperturbative quasi-particle interactions on top of the IDD effects, termed ``induced density dependence"  or DD$_{\rm induced}$  in \cite{CND-III}-- manifesting scale-chiral symmetry.

The result is plotted in Fig.~\ref{mu_L_PKLR}.  We see that $\mu_\Lambda-m_\Lambda$ crosses zero at a density $1.5 < n/n_0 < 2.0 $. The result is insensitive to the demarcation density for the regions. In fact what comes out in the mean field in $bs$HLS\footnote{We recall that here ``mean field" endowed with IDDs goes beyond what is used in the standard mean-field theory.}  is quite easy to understand. Since $m^\ast_\Lambda$ stops dropping with $f^\ast_\sigma$ stabilizing at $2n_0$, what matters is the interplay of the ratio $(\frac{g^\ast}{m^\ast})^2$ for the scalar and vector mesons -- with an opposite sign -- multiplied, respectively, by the scalar density $n_s$ and by the baryon number density $n$. The vector repulsion wins over the scalar attraction as density increases in the same way as it does in nuclear matter. Although the estimate is admittedly approximate -- and it could be done much more realistically in the $V_{lowk}$ approach used above,  that the BS constraint~\cite{bedaque-steiner} is met is most likely robust. We conclude from this that within the formalism developed in \cite{LPR:PR15} and with the prescription given in \cite{bedaque-steiner}, the hyperon problem could be avoided.  A more definitive answer awaits a full $V_{lowk}^{SU(3)}$RG calculation.

An intriguing -- and perhaps important -- point to note here is that the  density involved corresponds more or less to the skyrmion-half-skyrmion transition at $n_{1/2}\sim 2n_0$~\cite{bsHLS,CND-III} which corresponds roughly to the smooth hadron-quark (or quarkyonic) transition in the model where quark degrees of freedom (including strange quark) are explicitly incorporated~\cite{kojoetal}.

\section{Kaon condensation problem}

The mechanism that the discussion of hyperons vs. kaons given  above relied on was the binding of $K^-$'s to dense baryonic matter.  There in the presence of beta equilibrium, phase transitions were not involved. Unlike hyperons,  however,  kaons can bose-condense at high enough a density. In discussing the binding, what's involved in chiral Lagrangians is the leading chiral order (LO) term known as Weinberg-Tomozawa (WT) term. In the literature discussing on $\Lambda (1405)$ with respect to KN interactions, both free and bound in matter, it is this leading-order (LO) term  that plays the key role. This term however turns out to be marginally irrelevant from the RG point of view. More on this matter below. What is relevant  involves a scalar meson, i.e., the dilaton $\sigma$, the pseudo-NG boson from spontaneously broken scale symmetry discussed in \cite{bsHLS}. It is found that kaons condense ultimately after the appearance of hyperons when the dilaton condensate $\la\chi\ra$ goes to zero, i.e., at the dilaton limit fixed point~\cite{bsHLS}

\subsection{Kaons in skyrmion crystal}\label{kaons-in-skyrmion}\index{kaons!in skyrmion crystal}
In the Callan-Klebanov model with which the issue of hyperons vs. kaons was discussed, the topological Wess-Zumino term played the dominant role. Viewed in terms of the $bs$HLS Lagrangian, it is roughly equivalent to the $\omega$ exchange term which when localized, corresponds to the Weinberg-Tomozawa term in S$\chi$PT.  But this interaction is not the relevant interaction at high density.  It is rather the scalar dilaton that plays the crucial role in dense matter at $n > n_{1/2}\sim 2n_0$.

Since we are primarily interested in qualitative rather than numerically accurate features, one can use  the Skyrme model (skyrmion$_\pi$ with Nambu-Goldstone bosons only).  Strangeness is involved, so we will deal with three flavors with the topological Wess-Zumino term included.  The dilaton will be introduced in the way described in \cite{bsHLS,Li-Ma-Rho,CND-III}. Other heavy-meson fields, while quantitatively relevant, are not expected to modify the general features, so they will be ignored. The price to pay for this simplification is that the density at which the skyrmion-half-skrymion transition, which here again will figure importantly, takes place will not be given reliably.  I believe this does not affect the main argument.

%
Take the scale-chiral Lagrangian valid in the leading order in scale symmetry (``LOSS") as derived in \cite{CND-III,Li-Ma} and given a support in \cite{bsHLS, gA},
\begin{eqnarray}
{\cal L}_{sk}
&=& \frac{f^2}{4} \left(\frac{\chi}{f_\sigma}\right)^2
{\rm Tr} (L_\mu L^\mu) + \frac{1}{32e^2}{\rm Tr} [L_\mu, L_\nu]^2\nonumber\\
&& +\frac{f^2}{4}\left(\frac{\chi}{f_\sigma}\right)^3
{\rm Tr}{\cal M} (U+U^\dagger-2)
\nonumber \\
&&+\frac{1}{2}\partial_\mu \chi\partial^\mu \chi + V(\chi)\label{SK}
\label{lag1}
\end{eqnarray}
where $V(\chi)$ is the dilaton potential, $ L_{\mu} = U^{\dagger} \partial_{\mu} U $, with
$U$ the chiral field taking values in $SU(3)$. One ignores the pion mass for simplicity, so the explicit chiral symmetry-breaking mass term is given by the mass matrix ${\cal M} = {\rm diag} \left( 0, 0, 2m_{K}^2\right)$. In $SU(3)_f$, the anomaly term, i.e., the Wess-Zumino term,
${\cal S}_{WZ} = - \frac{i N_C}{240 \pi^2} \int d^5 x
\varepsilon^{\mu\nu\lambda\rho\sigma}
\mbox{Tr} \left( L_{\mu} L_{\nu} L_{\lambda} L_{\rho} L_{\sigma} \right)$,
turns out to play a crucial role in our approach:

Now consider the fluctuation of kaons in the background of the skyrmion matter $u_0$
\begin{equation}
  U(\vec x,t) = \sqrt{U_K(\vec x,t)} U_0(\vec x) \sqrt{U_K(\vec x,t)},\label{CKansatz}
\end{equation}
\begin{equation}
  U_K(\vec x,t) =
  e^{\frac{i}{\sqrt{2}f_{\pi}}
\left( \begin{array}{cc}
0 & K \\ K^\dagger &
0 \end{array} \right)},\
  U_0(\vec x)
    = \left( \begin{array}{cc} u_0(\vec x) & 0 \\ 0 & 1 \end{array} \right).
\end{equation}
Substituting (\ref{CKansatz}) into (\ref{lag1}) and the Wess-Zumino term,
one gets the kaon Lagrangian in the background matter field $u_0(x)$\footnote{Recall that this background-dependence  is supposed to effectively mock up the IDDs.} 
\begin{eqnarray}
{\cal L}_{K}
&=&
\bigg(\frac{\chi_0}{f_\sigma} \bigg)^2 \dot K^{\dagger}G\dot K
- \bigg(\frac{\chi_0}{f_\sigma} \bigg)^2 \partial_iK^{\dagger}G\partial_iK\nonumber\\
&& \hskip 2em - \bigg(\frac{\chi_0}{f_\sigma} \bigg)^3 m_{K}^2 K^{\dagger} K
\nonumber\\
&& \hskip 2em
 + \frac{1}{4} \bigg(\frac{\chi_0}{f_\sigma} \bigg)^2
   \left( \partial_\mu K^{\dagger} V^{\mu}(\vec x) K
 - K^{\dagger} V_\mu(\vec x) \partial^{\mu}K \right),
\nonumber \\
&&  \hskip 2em
+ \frac{iN_c}{4f_{\pi}^2} B^0
 \left( K^{\dagger}G\dot K - \dot K^{\dagger} GK \right).
\end{eqnarray}
where $\chi_0(\vec{x})$ is the classical dilaton field and
\begin{eqnarray}
V_\mu(\vec{x})
&=& \frac{i}{2} [(\partial_\mu u_0^\dagger)u_0 - (\partial_\mu u_0) u_0^\dagger],
\label{V} \\
G(\vec x) &=& \frac{1}{4} (u_0+u_0^{\dagger}+2),
\\
B^{\mu}(\vec x) &=& \frac{1}{24\pi^2} \varepsilon^{\mu\nu\lambda\sigma}
 {\rm Tr}\left( u_0^{\dagger} \partial_{\nu} u_0
 u_0^{\dagger} \partial_{\lambda} u_0 u_0^{\dagger}
 \partial_{\sigma} u_0 \right).
\end{eqnarray}
In the spirit of mean field approximation, the space average is taken on the
background matter fields $u_0$ and $\chi_0$,  obtaining
\begin{equation}
{\cal L}_{K}
= \alpha ( \partial_\mu K^{\dagger} \partial^\mu K)
+ i \beta ( K^{\dagger} \dot K - \dot K^{\dagger} K )
- \gamma K^{\dagger} K
\label{MFA}
\end{equation}
where
\begin{eqnarray}
\alpha = \left< \kappa^2 G \right>,
\beta = \frac{N_c}{4f_{\pi}^2} \left< B^0 G \right>,
\gamma = \left<\kappa^3 G \right> m_{K}^2
\end{eqnarray}
with $\kappa=\chi_0/f_\chi$.
Lagrangian (\ref{MFA}) yields a dispersion relation for the kaon in the
skyrmion matter as
\begin{equation}
\alpha( \omega_K^2 - p_K^2) + 2 \beta \omega_K + \gamma = 0.
\end{equation}
Solving this for $\omega_K$ and taking the limit of $p_K \rightarrow 0$,
one gets 
\begin{equation}
m_K^* \equiv \lim_{p_K \rightarrow 0} \omega_K
= \frac{-\beta + \sqrt{\beta^2 + \alpha\gamma}}{\alpha}.
\end{equation}
This equation will be used for evaluating the in-medium effective kaon mass.

The key element in Eq.~(\ref{MFA}) for the kaonic fluctuation is the background $u_0$ which reflects the ``vacuum" modified by the dense skyrmion matter. The classical dilaton field tracks the quark condensate affected by this skyrmion background $u_0$ that carries information on chiral symmetry of dense medium. Our approach here is to describe this background $u_0$ in terms of crystal configuration. The pertinent $u_0$ has been worked out in detail in \cite{half-skyrmions}, from which we shall simply import the results for this work. As shown there, skyrmions put on an FCC -- which is the favored crystal configuration -- make a  transition at a density $n_{1/2}$ to a matter consisting of half-skrymions.
With the parameters of the Lagrangian picked for the model, we find $n_{1/2}\sim 1.3 n_0$ but this is not significant. As noted before, the reasonable range of density  range  is $n_{1/2}\sim (2-3) n_0$.

\begin{figure}
\centerline{\includegraphics[width=9.0cm]{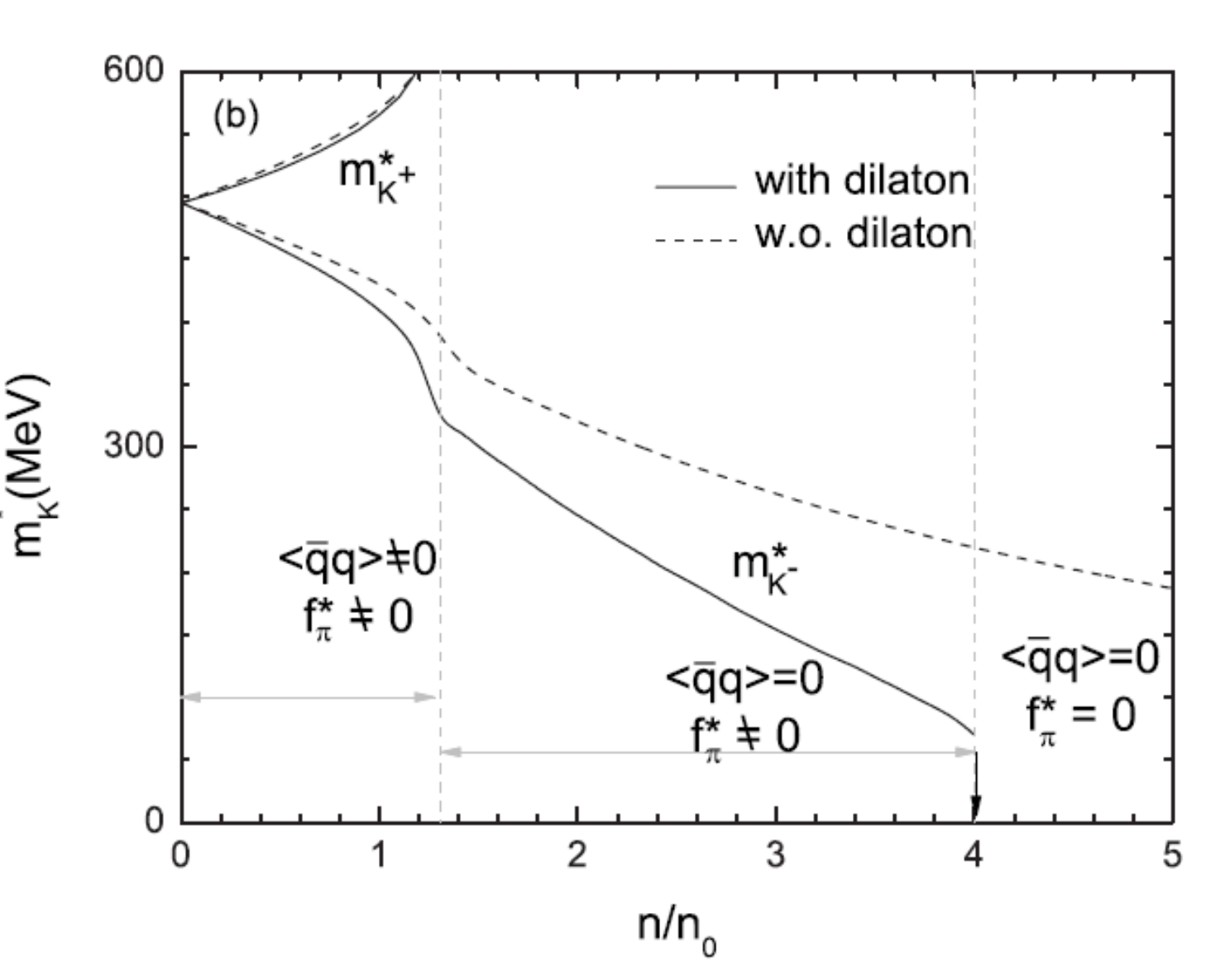}}
\caption{ $m^*_{K^\pm}$ vs. $n/n_0$ (where $n_0\simeq 0.16$ fm$^{-3}$ is the normal nuclear matter density) in dense skyrmion matter which consists of three phases: (a) $\la\bar{q}q\ra\neq 0$ and $f_\pi^*\neq 0$, (b) $\la\bar{q}q\ra=0$ and $f_\pi^*\neq 0$ and (c) $\la\bar{q}q\ra=0$ and $f_\pi^*=0$. The parameters are fixed at $\sqrt{2}ef_\pi=m_\rho=780$ MeV and dilaton mass $m_\sigma=720$ MeV. }
\label{kaon-in-crystal}
\end{figure}

We now discuss the results given in Fig.~\ref{kaon-in-crystal} in some detail.

The results are given for the dilaton mass $m_\sigma\approx 720$ MeV. The dilaton mass is currently known  neither experimentally nor theoretically. It is taken as $m_\sigma \approx 720$  so as to give $\sim 600$ MeV in nuclear matter consistent with the Fermi-liquid description of nuclear matter implementing IDDs~\cite{song-brown-min-rho}. The density at which the half-skyrmion matter appears turns out to be more or less independent of the dilaton mass~\cite{half-skyrmions}. In fact, the $n_{1/2}$ is entirely dictated by the parameters that give the background $u_0$  (for which the hidden local symmetric mesons cannot be ignored). For the skyrmion background, $\sqrt{2}ef_\pi$ -- which is the vacuum $\rho$ mass at tree order-- can be taken to be $\sim 780$ MeV which fixes $e$ for the given $f_\pi\approx 93$ MeV. What is controlled by the dilaton mass are the density at which the dialton condensate vanishes together with the pion decay constant, i.e., the dilaton limit fixed point and more significantly, the rate at which the kaon mass drops as density exceeds $n_{1/2}$.

What is novel in this model is that the behavior of the kaon is characterized by three different phases, not two as in conventional approaches. In the low density regime up to $\sim n_{1/2}$, the kaon interaction can be described by standard chiral symmetry treatments. For instance, the binding energy of the kaon at nuclear matter density comes out to be $\sim (80-100)$ MeV for $m_\sigma= 720$ MeV. Going above $n_{1/2}$, however, as the matter enters the half-skyrmion phase with vanishing quark condensate (on average) and non-zero pion decay constant, the kaon mass starts dropping more steeply. This form of matter arising from the skyrmion-half-skyrmion transition involving the dialton condensate -- which is undoubtedly highly nonperturbative missing in S$\chi$PT  -- is  not captured by the Weinberg-Tomozawa term. Finally at $n_c$ at which both the quark condensate and the pion decay constant vanish, the kaon mass vanishes. This corresponds to the dilaton limit fixed point mentioned above. This happens in this model at density $\approx 4.0 n_0$ -- which is clearly much too low -- for the dilaton mass $m_\sigma=720$ MeV. The reason for this is that the dilaton condensate, $\la\chi\ra$, vanishes at the dilaton limit fixed point. This means that ``kaon condensation" in symmetric nuclear matter  takes place at the point at which the scale symmetry associated with the soft dilaton is restored. In the realistic treatment given in \cite{bsHLS}, it was found that the vector manifestation fixed point and the dilaton limit fixed point are to arrive at $n >\sim 10 n_0$. Therefore this anomalous behavior of the kaon mass at the skyrmion-1/2-skyrmion transition must take place at a much higher density than what's estimated in the simple model given above. In fact this observation will be supported in the renormalization-group treatment given below.

\subsection{Kaons in Fermi liquid}\index{kaons!in Fermi liquid}
The crucial role of the dilaton in the strangeness problem in dense compact stars  could be unravelled in a $V_{lowk}^{SU(3)}$RG treatment. In the absence of such a treatment, we look at a greatly simplified model that captures essential ingredients involved in the process. 

The model is formulated as follows.  Given that the $V_{lowk}$RG in two flavors employed in \cite{bsHLS}  is equivalent  to doing Landau-Fermi liquid theory with a $bs$HLS Lagrangian endowed with IDDs as argued in \cite{holt-fermiliquid}, the formalism is extended to three flavors.  The key requirement is to account for both nuclear interactions in the Fermi liquid and kaonic fluctuations on top in Wilsonian RG, taking into account kaon properties warped by the baryonic background in dense medium as indicated by the result of Pandharipande et al.~\cite{pandharipande-thorsson}. There are in principle many  terms in the action. The simplification consists of  localizing  $V_{lowk}$s in all three channels, namely baryon-baryon, baryon-kaon and kaon-kaon.   Assuming that the process takes place in the half-skyrmion phase from which kaons are to condense is in Fermi liquid, we then look at the kaons coupling to the Fermi liquid via exchange of a scalar dilaton that can be tuned to criticality (i.e., zero mass). We ignore the effect of the electron chemical potential~\cite{BTKR} in this discussion.
We write the simplified Euclidean action for kaon-Fermi-liquid nuclear systems in momentum space in the form~\cite{PR-fermiliquid}
\begin{equation}
Z = \int [d\phi][d\phi^\ast][d\psi][d\bar{\psi}] { e}^{-\tilde{S}^E}
\end{equation}
with the actions
\begin{eqnarray}
\tilde{S}^E&=& \tilde{S}_\psi^E + \tilde{S}_\phi^E +\tilde{S}_{\psi\phi}^E\ \label{S2}\\
\tilde{S}_\psi^E &=&\int  \frac{d\epsilon  d^3 \vec{k}}{(2\pi)^4}
\,\bar{\psi}_\sigma \left\{ -i\epsilon +(e(\vec{k})-e_F)\right\} \psi_\sigma\nonumber\\
&& - \int \left( \frac{d\epsilon  d^3 \vec{k}}{(2\pi)^4}\right)^4 \lambda \bar{\psi}_\sigma \bar{\psi}_{\sigma^\prime} \psi_\sigma\psi_{\sigma^\prime} \delta^4(\epsilon,\,\vec{k})\label{fermion2}\\
\tilde{S}_\phi^E &=& \int \frac{d\omega d^3\vec{q}}{(2\pi)^4} \{\phi^*(\omega^2 +q^2)\phi +m_\phi^2\phi^*\phi+\cdots\} \label{boson2}\\
\tilde{S}_{\psi\phi}^E &=&-\int \left(\frac{d\epsilon d^3\vec{k}}{(2\pi)^4} \right)^2 \left(\frac{d\omega d^3 \vec{q}}{(2\pi)^4} \right)^2 \, h\,\phi^*\phi \bar{\psi}_\sigma\psi_\sigma \delta^4(\omega,\,\epsilon,\,\vec{q},\,\vec{k})\,.\nonumber\\
\label{couple2}
\end{eqnarray}
where $\psi$ and $\bar{\psi}$ are the eigenvalues of $\Psi$ and $\Psi^\dagger$ acting on $\left|\psi \right. \rangle$ and $\langle \left. \bar{\psi} \right|$, fermion coherent state,
\begin{equation}
\Psi \left|\psi \right. \rangle = \psi \left|\psi \right. \rangle \quad  {\rm and} \quad \langle \left. \bar{\psi} \right| \Psi^\dagger = \langle \left. \bar{\psi} \right| \bar{\psi}\,.
\end{equation}
Although our notations for the fields are general as they can apply to other systems like pions/nucleons and electrons/phonons, we should keep in mind that we are specializing to the $K^-$ field for the boson and the non-strange baryons, proton and neutron for the fermion.\footnote{An action similar to (\ref{S2}) was is  studied in \cite{LRS} except for the bosonic action. The bosonic action used there is different from (\ref{boson2}) in that the system considered there was assumed to be a compact-star matter in weak and chemical equilibrium (with the boson approximated to satisfy a quadratic dispersion relation  is most likely incorrect for kaons) whereas here we are dealing with relativistic bosons which could be tuned to criticality by attractive KN interactions.}  It is interesting that a same-type action figures in condensed matter physics. In fact, (\ref{S2}) is quite similar to the action studied for quantum critical metals in \cite{kachru} from which we shall borrow various scaling properties for the renormalization group.

We have taken here a local KN interaction which in terms of our $bs$HLS Lagrangian can be interpreted as arising from the exchanges of the dilaton and $\omega$ meson with the mesons integrated out.  Similarly we have taken only one generic four-Fermi interaction. Hyperons could figure in the Fermion sector but we will focus on nucleons. Therefore huperon problem is not addressed directly. More realistically it would have various channels, including the channels corresponding to the dilaton exchange and the $\omega$ exchange. The intricate interplay between the various channels, in particular the scalar and vector channels, must play a crucial role in giving the mechanism leading to  kaon condensation. One notable observation is that in the nucleon-nucleon interaction, the attractive scalar exchange and repulsive vector exchange compete whereas in the KN interaction (where N is a neutron) both are attractive and add.  We will not go into these intricacies here but just give a generic feature of what a consistent treatment might mean in a simple situation. Needless to say, this simple treatment cannot possibly be truly realistic.

It is instructive  to compare the constant $h$ in (\ref{couple2}) with what is given in the standard chiral Lagrangian.  In terms of the chiral counting, the leading term ($O(p)$) is ``Weinberg-Tomozawa-type" term, coming from the exchange of the $\omega$ meson  and gives, for the s-wave kaon,
\be
h_{WT}\propto q_0/f_\pi^2
\ee
where $q_0$ is the fourth component of the kaon 4-momentum
and the next chiral order ($O (p^2)$) term is the ``KN sigma-type" term
\be
h_{\Sigma}\propto \Sigma_{KN}/f_\pi^2
\ee
which can be thought of arising from a scalar meson exchange. In our approach, both terms can be considered as resulting from integrating out the ``heavy" mesons from the $bs$HLS Lagrangian.

We should stress that the fermion $\psi$ is a quasiparticle field in Fermi-liquid near its fixed point, so the coupling $h$ encodes more than just the ``bare" couplings in the $bs$HLS  Lagrangian. This point was made in \cite{MR-prl91}.  For convenience, we shall use these chiral Lagrangian terms for identifying the two terms contributing to $h$. It should be remembered that they are not the same.

As written,  the action for the fermion system (\ref{fermion2}) is for a Fermi liquid with marginal four-Fermi interactions~\cite{shankar}. So the energy-momentum of the quasiparticle (fermion) is measured with respect to the Fermi energy $e_F$ and Fermi momentum $k_F$. The bosonic action is for massive Klein-Gordon field, which later will be associated with a pseudo-Goldstone field, with higher-field terms ignored.  In contrast to that of the quasi-particle, the boson energy-momentum will be measured from zero.  We would like to look at meson-field fluctuations on the background of a fermionic matter given by the Fermi-liquid theory. The Fermi-liquid structure is expected to be valid as long as $N\equiv k_F/\bar{\Lambda} \gg 1$ where $\bar{\Lambda}=\Lambda-k_F$ where $\Lambda$ is the (momentum) cutoff scale from which mode decimation -- in the sense of Wilsonian -- is made.  For nucleon systems, the action (\ref{fermion2}) gives the nuclear matter stabilized at the fixed point, Landau Fermi-liquid fixed point, with corrections suppressed by $1/N$~\cite{shankar}. The four-Fermi interactions and the effective mass of the quasiparticle are the fixed-point parameters. We expect that as long as $N\gg 1$, the Fermi-liquid action can be trusted even at higher densities than $n_0$. Recall that the skyrmion-half-skyrmion ``transition" which figures importantly is not a phase transition in the sense of Ginzburg-Landau-Wilson paradigm. This is because the half-skyrmions are confined\footnote{That half-skyrmions are confined in the strong interactions via monopoles is suggested in \cite{confined halfskyrmions}.} although their presence makes a dramatic impact on the EoS.

For simplicity, we shall assume a spherically symmetric Fermi surface in which case we can set for the quasiparticle momentum
\be
\vec{k}=\vec{k}_F+\vec{l}\approx \hat{\Omega}(k_F+l).\label{fermionmomentum}
\ee
Then for $\bar{\Lambda} \ll k_F$, we have
\begin{equation}
e(k)-e_F \approx \vec{v}_F \cdot \vec{l} +O(l^2)
\end{equation}
where $v_F$ is the Fermi velocity $v_F=k_F/m_L^\star$ with $m_L^\star$ the effective quasiparticle (Landau) mass.
\subsubsection{Scaling}
To do the Wilsonian decimation, we need to set {the cutoff scale ${\Lambda}$} at which the classical action has bare parameters. Quantum effects are taken into account  by doing the mode elimination  by lowering {the cutoff from ${\Lambda}$ to $s{\Lambda}$} with $s<1$.  The important point to note here is that the boson and fermion fields have different kinematics. While the boson momentum is measured from the origin -- and hence the momentum cutoff is $\Lambda$ , the fermion momentum is measured from the Fermi momentum $k_F$. Therefore the mode elimination for the fermion involves lowering the fermion momentum from $\bar{\Lambda}=\Lambda-k_F$ to $s\bar{\Lambda}$. As noted above, the strategy in the fermion sector is to take the large $N$ limit where $N\equiv k_F/\bar{\Lambda}$.\footnote{We remind the readers  that the larger the $k_F$ or density, the more $\bar{\Lambda}$ shrinks, which would mean that the large $N$ argument would hold better in the fermion sector as the density increases. This suggests that the mean field approximation with effective Lagrangians -- chiral Lagrangian or hidden local symmetry Lagrangian -- would get better the higher the density. }


 We define the scaling laws of fields and other quantities like the delta functions by requiring that the kinetic energy terms of the fermion and the boson be invariant under the scaling
\be
\epsilon\rightarrow s\epsilon, \ \omega\rightarrow s\omega, \ l \rightarrow sl, \ \vec{q}\rightarrow s\vec{q}. \label{vector}
\ee
Only the fermion momentum orthogonal to {the Fermi surface} scales in the way the fermion energy and all components of the boson momentum do. Since the quasiparticle mass {$m_L^\star$}  is a fixed-point quantity, the Fermi velocity does not scale. Therefore we have the fields scaling as
\be
\phi&\rightarrow& s^{-3}\phi,\nonumber\\
\psi&\rightarrow& s^{-3/2}\psi\,, \label{field}
\ee for which we denote the scaling dimensions as $[\phi]=-3$ and $[\psi]=-3/2$\footnote{This notation will be used in what follows.} and the meson mass term scales
\begin{equation}
\int d\omega d^3q\, m_\phi^2 \phi^\ast \phi \rightarrow \int d\omega d^3q\, m_\phi^2 s^{-2} \phi^\ast \phi = \int d\omega d^3q\, m_\phi^{\prime\,2} \phi^\ast \phi,
\end{equation} with $m^\prime_\phi \equiv s^{-1} m_\phi$. This shows the well-known fact that the meson mass term is ``relevant."
Using the procedure of scaling toward the Fermi surface, we have
\be
[d\epsilon d^3k]=2,\  [\delta (\epsilon,k)] = -2\,. \label{fdelta}
\ee
This confirms that  the four-Fermi interaction term in (\ref{fermion2}) is marginal.\footnote{For two or three spatial dimensions, this applies to only forward scattering and BCS-type scattering. Others are marginal.}

The scaling of the coupling term (\ref{couple2}) is more subtle. Using the scaling dimensions
\be
[\phi]=-3, \ [\psi]=-3/2, \ [d\epsilon d^3\vec{k}]=2, \ [d\omega d^3\vec{q}]=4,\label{scaling-couple}
\ee
we find the scaling of the integrand $I_{\psi\phi}$ of the action (\ref{couple2})  written as $\int I_{\psi\phi}$ is
\be
[I_{\psi\phi}]=3+[h]+[\delta (\omega,\epsilon,\vec{q},\vec{k})],\label{Iscaling}
\ee
where the bare coupling constant $h$ will have the scaling dimension $[h_{WT}]=[q_0]=1$ and $[h_{\Sigma}]=0$. The $\psi\phi$ coupling will be ``relevant" if  $[I_{\psi\phi}]<0$, marginal if $=0$ and ``irrelevant" if $>0$. It is thus the scaling of the delta function that determines the scaling of the coupling $h$.

It will follow from arguments based on $bs$HLS developed so far that the $\delta$ function scales as
\be
[\delta (\omega,\epsilon,\vec{q},\vec{k})]=-4. \label{deltascaling}
\ee
This then leads to
\be
[I_{\psi\phi}]_{h_{WT}}=0, \  [I_{\psi\phi}]_{h_{\Sigma}}=-1.\label{II}
\ee
Thus the Weinberg-Tomozawa (WT)-type coupling is marginal,
while the sigma-term-type coupling is relevant. The one-loop contrition to the former turns out to be irrelevant\cite{PR-fermiliquid}, so the WT-type term is marginally irrelevant.Therefore one can focus  on the latter.

Briefly the scaling (\ref{deltascaling}) is arrived at as follows. (For details, we refer to \cite{PR-fermiliquid}.)

First note that the KN coupling giving rise to (\ref{couple2}) differs from that leading to the four-Fermi interaction (\ref{fermion2})  by that one of the two $\psi\psi\phi$ vertices in the former is replaced by a $KK\phi$ vertex. Within the scaling rule we adopt, the $\psi\psi\phi$ vertex is marginal but the $KK\phi$ vertex is relevant. All other quantities are the same. Since in our approach, we want the four-Fermi interaction to remain marginal
so that the baryonic matter remains in Fermi liquid, we need to enforce that the presence of the $\phi$ field leave unaffected the marginal four-Fermi interaction. This is achieved by assigning a definite scaling property to the $\phi$ propagator. By imposing this scaling condition on the $\phi$ propagator that figures in the KN-KN interaction, one sees immediately that the $\Sigma$-term type KN coupling in (\ref{couple2}) must be relevant. This is simply because the $KK\phi$ vertex is relevant. One can verify this result by an explicit counting of the scalings involved in the Feynman diagrams~\cite{PR-fermiliquid}.

The same simple counting rule shows that  the WT term should be marginal. For this, simply replace the $\psi\psi\phi$ vertex of the scalar-exchange term by a $\psi\psi\omega$ coupling and the $KK\phi$ vertex by a $q_0 KK \omega$ vertex, which amounts to replacing a marginal coupling by a marginal coupling and a relevant coupling by a marginal coupling. The net result is marginal because $q_0 KK\omega$ vertex is marginal.
\subsubsection{Wilsonian renormalization group}
We are interested in the flows of the kaon mass and the kaon-nucleon coupling and the effect of kaon condensation on the Fermi liquid structure of the baryonic system. For this we need the scaling properties of the diagrams. At tree order, we have
\begin{equation}
h^\prime = s^{-1}\,h\,,
\end{equation}
i.e., relevant as found above. The kaon mass scaling is of course relevant,
\be
m_\phi^{\prime\, 2}=s^{-2} m_\phi^2.
\ee
\begin{figure}[ht]
\begin{center}
\includegraphics[width=6.0cm]{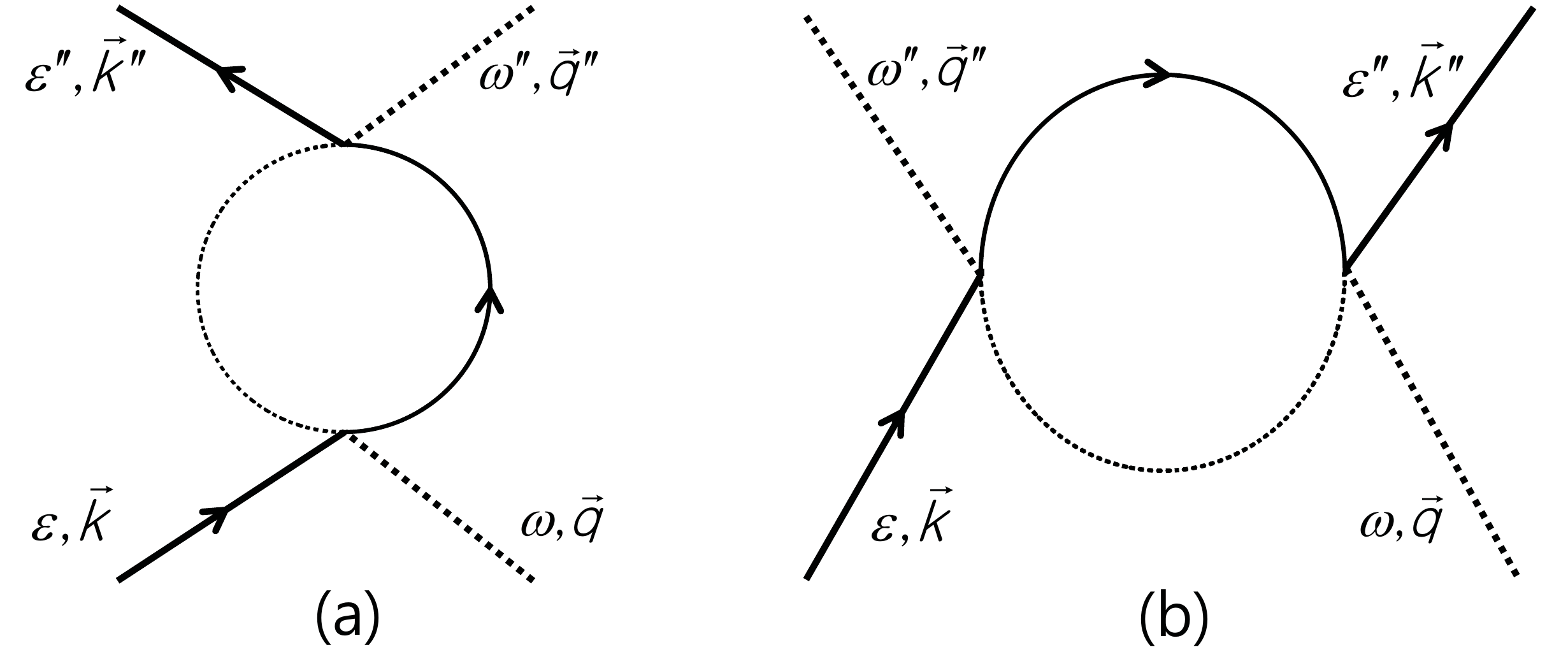}
\caption{One-loop graphs contributing to $h$. The process corresponds to f(ermion) ($\epsilon,\vec{k}$)+ m(eson) ($\omega,\vec{q}$)$\rightarrow$  f(ermion) ($\epsilon^{\prime\prime},\vec{k}^{\prime\prime}$)+ m(eson) ($\omega^{\prime\prime},\vec{q}^{\prime\prime}$).}
\label{kaon_fermion}
\end{center}
\end{figure}
\begin{figure}[bh]
\begin{center}
\includegraphics[width=3.0cm]{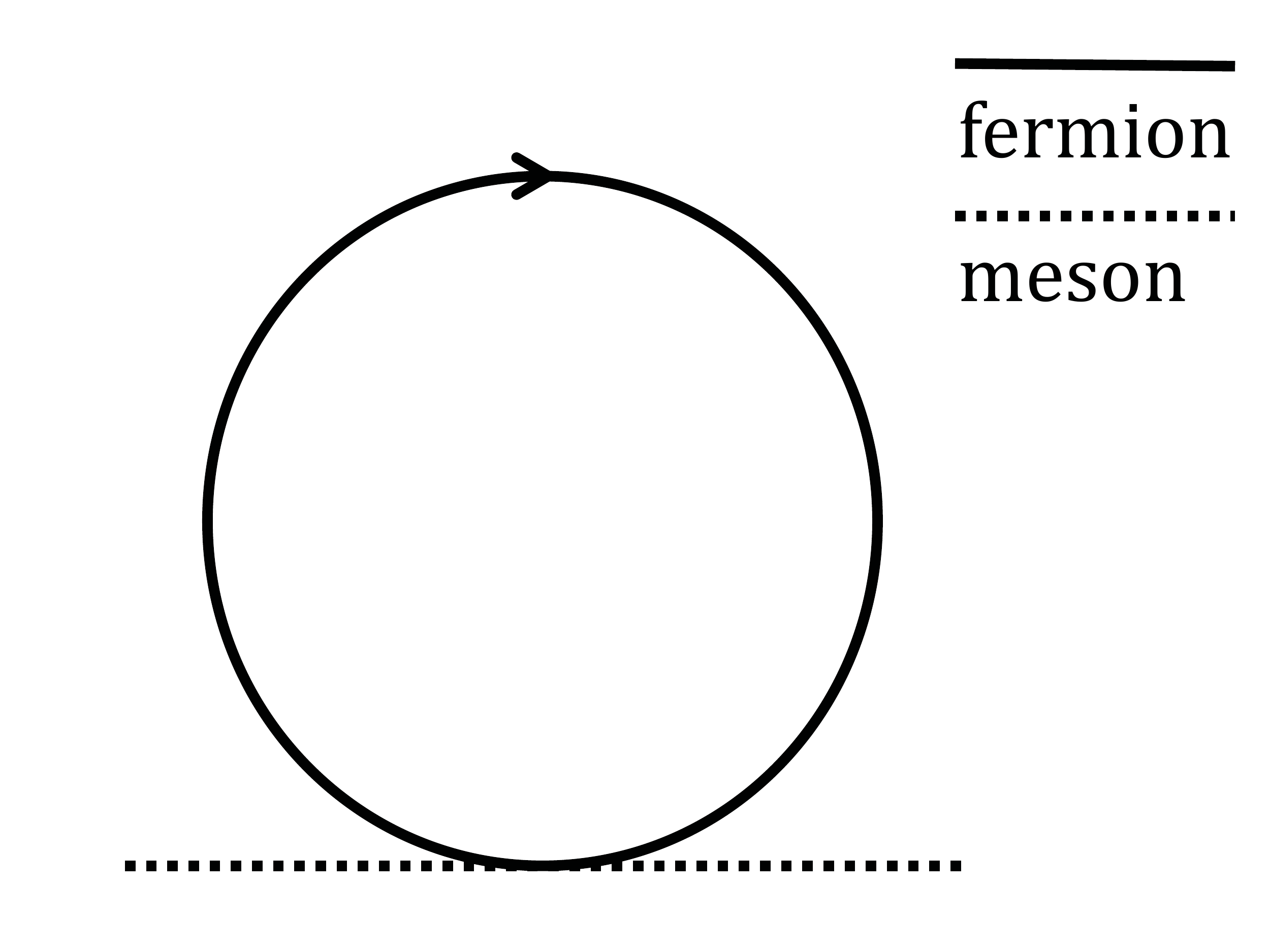}
\caption{One-loop graph contributing to $m_\phi$.}
\label{one_loop_mass}
\end{center}
\end{figure}

At one-loop order, there are two diagrams, Figs.~\ref{kaon_fermion}, contributing to $h$ and one diagram, Fig.~\ref{one_loop_mass}, to $m_\phi$.
%
%
The one-loop graph Fig.~\ref{one_loop_mass} contributing to the $\phi$ mass is easy to evaluate. By decimating from $\bar{\Lambda}$ to $s\bar{\Lambda}$ in the nucleon loop, we get
{\begin{eqnarray}
\int \frac{d \omega d^3\vec{q}}{(2\pi)^4} \phi^\ast \phi s^{-2}\left[ -m_\phi^2 +\frac{\gamma h k_F^2 }{2\pi^2} \int_{s\bar{\Lambda} < \left|l^\prime \right| < \bar{\Lambda}} dl^\prime\, \left|{\rm sgn}\left( l^\prime \right) \right|\right]\,,\nonumber\\
\end{eqnarray}
where $\gamma$ is the degeneracy factor for flavor (=2) and spin (=2): $\gamma=4$ for nuclear matter and $\gamma=2$ for neutron matter. It follows from above that
\begin{equation}
\delta \left[m_\phi^2 \right] = s^{-2} \left[ m_\phi^2 - \frac{\gamma h \bar{\Lambda} k_F^2 }{\pi^2} \left( 1-s \right) \right] - m_\phi^2\,.
\end{equation}}

As for the loop contribution to the $h\phi\phi\psi\psi$ coupling, it turns out that the two diagrams (a) and (b) of Fig.~\ref{kaon_fermion} are related to each other. And they are down by $1/N$ relative to the tree term. Furthermore both are found to vanish~\cite{PR-fermiliquid}. Therefore to one-loop order the matter becomes very simple.

\subsubsection{Flow analysis}

From what's discussed above, it is straightforward to write down the RGEs. Setting $t\equiv -\ln s$,
\begin{eqnarray}
\frac{d\, m_\phi^2}{dt} =  2\, m_\phi^2 - A \, h\,, \label{massrge}\\
\frac{d\, h }{dt} =  h - B\, h^2\,, \label{hrge}
\end{eqnarray} where{
\begin{eqnarray}
A &=& \frac{\gamma k_F^2 \bar{\Lambda}}{\pi^2}=\frac{\gamma k_F^3}{N\pi^2}\,, \\
B &=& 0 \,.
\end{eqnarray}}
It is easy to get the analytic solutions for $m_\phi$ and $h$
\begin{eqnarray}
m_\phi^2 (t) &=& \left( m^2_\phi(0) - h(0) A \right) {\rm e}^{2t} + h(0) A{\rm e}^{t}\,,\label{mphi}\\
h(t) &=& h(0) {\rm e}^{t}\,.\label{hterm}
\end{eqnarray}

Given the analytic solution, we can work out how $m_\phi^2$ and $h$ flow as $t$ increases ($s$ decreases) for given values of $A$, $h(0)$ and $m_\phi^2(0)$.
From Eqs.~ (\ref{mphi}) and (\ref{hterm}), we have the relation,
\begin{equation}
\frac{m_\phi^2(t) - Ah(t)}{\left[ h(t) \right]^{2}} = \frac{m_\phi^2(0) - A\,h(0)}{\left[ h(0) \right]^{2}}\,, \label{rel_c1}
\end{equation} which is satisfied for any value of $t$. It is convenient to introduce the new parameter $c_{a_h}$ which is independent of $t$,
\begin{equation}
c\equiv \frac{m_\phi^2(0) - A\,h(0)}{\left[ h(0) \right]^{2}}\,. \label{rel_c2}
\end{equation}
The quantities at $t=0$,  i.e., $m_\phi^2(0)$ and $h(0)$, are given at $s=1$, that is, at the scale from which the decimation starts for a given $k_F$. {\it These parameters in the bare action -- and the parameter $c$ -- depend only on density. Since the flow depends on $c$, the RG properties of dense medium depend on the density dependence of the parameters of the bare chiral Lagrangian at the scale  ${\Lambda}$. This is equivalent  to the IDD that is obtained in relativistic mean field treatment of effective Lagrangians of the HLS-type which is again equivalent to  Landau Fermi-liquid theory.}

\begin{figure}[ht]
\begin{center}
\includegraphics[width=8.0cm]{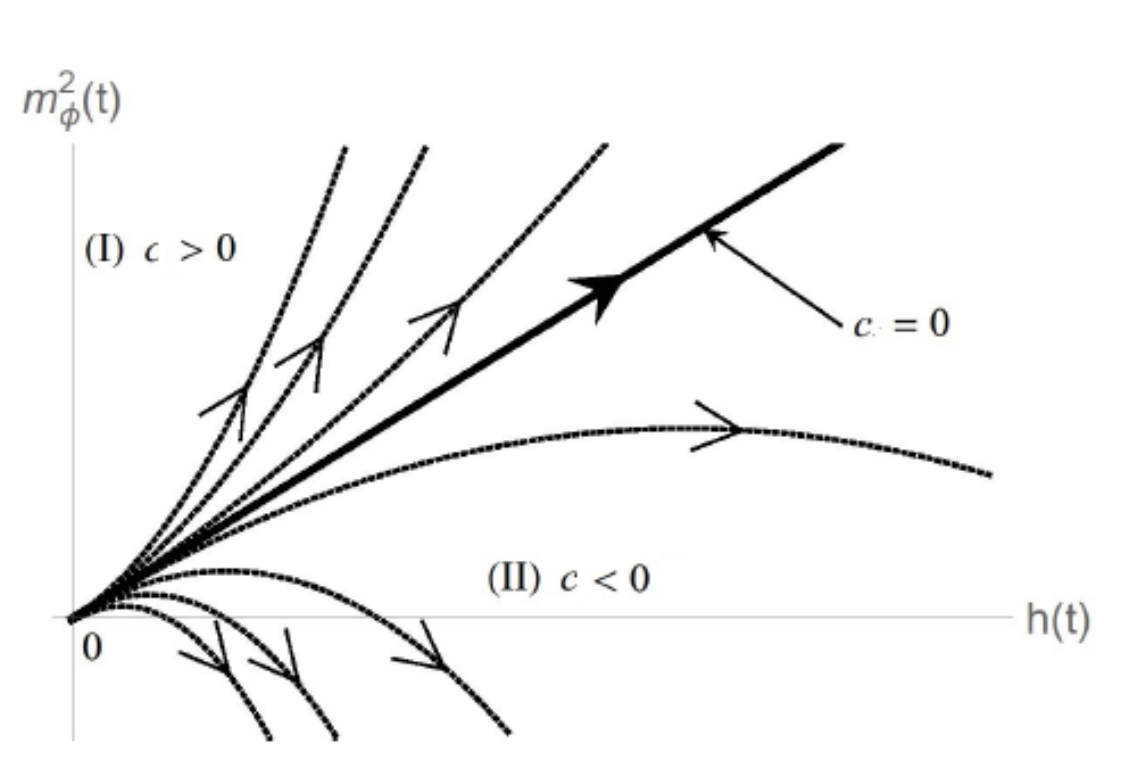}
\caption{RG flows of the parameters, $m_\phi^2(t)$ and $h(t)$, are shown in the parameter space. The lines are drawn with the fixed value of A and for different  $c$'s. The black solid line stands for the ``critical line (or surface)" that delineates the parameter spaces. }
\label{RG_flow}
\end{center}
\end{figure}
Now using Eqs.~(\ref{rel_c1}) and (\ref{rel_c2}), we readily obtain
\be
m_\phi^2(t)
&=&  c h(t)^2 + A\,h(t).
\ee
The flows $m_\phi (t)$ vs. $h(t)$  are plotted in Fig.~\ref{RG_flow}  for various values of $c_{a_h}$ that depend on $k_F/\bar{\Lambda}$. We note that $m_\phi^2(t)$ flows toward zero and becomes negative for $c < 0$. This signals the instability toward kaon condensation. What happens after the condensation is a matter that goes beyond the model. It could be stabilized by terms higher order in the kaon field not taken into account in our analysis.

The conclusion is that only the $m_\phi(0)$ and $h(0)$ that  satisfy
\be
m_\phi^2(0) &<& A h(0)\label{cond2kc}
\ee
will be in the parameter space for the condensation to take place. The value $c= 0$ defines the critical line in the $h(t)-m_\phi^2(t)$ plane, any point on which the parameters flow from  the origin satisfying
\be
m_\phi^2(t) &=& A\, h(t).
\ee
Note that in the RG flows of $m_\phi(t)$ and $h(t)$,  as $t$ goes up, the attractive interaction gets stronger.
\subsubsection{Instability toward the condensation}
Although one cannot determine the critical density at which kaons condense from the RG flow analysis,  it is nonetheless feasible to indicate whether and how the condensation can take place. This can be done by looking at the ``parameter window" that signals the instability toward condensation. 

At some density, according to the values of $c$, the RG flows of $m_\phi^2 $ and $h(t)$ will follow certain of the flow lines in Fig.~\ref{RG_flow}.
If the value of $c$ is changed by changing $k_F$ to $k_F^\prime$, the RG flow of $m_\phi^2$ and $h(t)$ follows from the line (in $k_F$) to the line (in $k_F^\prime$).
By locating the sign change of
\begin{equation}
m_\phi^2(0)- A\,h(0)\,,\label{kaonmass}
\end{equation}
one can then locate the density at which the phase is moved to the region that signals the instability toward the kaon condensation. While this argument does not allow to pinpoint what the condensation density can be, it nevertheless can give the {\it lower bound} of the critical density.  Let's use Eq.~(\ref{cond2kc}) to make this estimate. The sign change of $c$ takes place when
\begin{equation}
n\equiv \gamma\frac{k_F^3}{6\pi^2} = \frac{N}{6}\frac{m_\phi^2(0)}{h(0)}.\label{change}
\end{equation}
 {\it What this says is that since the better the quasiparticle picture is in the baryon sector which would be the case the larger $N$ is, the higher the critical density $n_K$. } This indicates that since the half-skyrmion phase, as argued,  is very likely in Fermi liquid at its fixed point~\cite{bsHLS}, if kaons do condense from the half-skyrmion phase which sets in at $n\gsim n_{1/2}\sim 2n_0$,  $n_K$ will inevitably be  high.  In compact stars in beta equilibrium,  the electron chemical potential $\mu_e$ that increases with density, would help reduce $n_K$ but the qualitative result obtained is not expected to be modified seriously. 
 
Just to have an idea of what sort of density range  is involved, let me take  $h(0)=\Sigma_{KN}/f^2$ from chiral Lagrangian. This is just to get a rough idea since the parameter $h$ should in our formulation represent coupling of a kaon on top of the Fermi sea to a quasiparticle warped by the background, both IDD and DD$_{induced}$, which of course could be quite different from $\Sigma_{KN}/f^2$.  From (\ref{change}), replacing $\phi$ by $K$ and $f$ by $f_\pi$, we find the condensation density  $n_K$ bound by
\be
n_K > \frac{N}{6}\frac{m_K^2 f_\pi^2}{\Sigma_{KN}}.
\ee
For numerical estimation, take $m_K \approx 500$ MeV and $\Sigma_{KN}\approx 250\,  {\rm MeV}$ using a recent lattice measurement for $\Sigma_{KN}$~\cite{sigmaterm}.  This gives at the Fermi-liquid fixed point
\be
n_K > N m_\pi^3/3\approx N n_0\to \infty.\label{infiniteN}
\ee
In order to have an idea what $N$ is, let's take the cutoff used in the $V_{lowk}$RG approach applied in \cite{bsHLS}. The optimal cutoff turns out to be $\Lambda\sim (2-3)$ fm$^{-1}$~\cite{ Vlowk}.  If one takes $\Lambda\sim 2$ fm$^{-1}$, then in the half-skyrmion phase that sets in at $n\gsim 2n_0$, $N=\frac{k_F}{\Lambda-k_F} \gsim 5 n_0$. This means once the half-skyrmion state sets in, the system rapidly moves towards the Fermi-liquid fixed point (with zero beta function). In the density range $(6-7) n_0$ relevant to the interior of massive compact stars, we will have $1/N\sim 0$. This implies that the condensation density  $n_K$ will be pushed way beyond the maximum interior density of massive stars.
\section{Conclusion}
Combining  the skyrmion description of dense matter with  possible smooth hadron matter-quark matter transition traded in for topology change at a density $n\sim (2-3)n_0$ and an Wilsonian RG analysis of coupled baryon-meson systems in dense medium,  I argued that (1) hyperons and condensed kaons must appear simultaneously in the large $N_c$ limit and (2) kaon condensation is banished to very high density beyond the density relevant to the interior of massive compact stars. The conclusion is that strangeness is unlikely to figure {\it explicitly} and cause the  strangeness (hyperons and condensed kaons)  problem in compact stars. 

I should mention a caveat here. The reasoning that $N$ is large in (\ref{infiniteN}) relies on that the matter treated is a good Fermi liquid. However the four-Fermi interaction corresponding to a scalar exchange in (\ref{couple2}), though a fairly good approximation near normal nuclear matter density~\cite{gelmini-ritzi}, cannot be valid  when the dilaton limit with the vanishing dlaton mass is approached. In other words, the Fermi-liquid structure must break down. What happens when this breakdown occurs is unknown. It is however most likely that the breakdown takes place at a  density much higher than is relevant for compact stars. A $V_{lowk}^{SU(3)}$ calculation could settle this issue.

\subsection*{Acknowledgments}
I am grateful for discussions with Hyun Kyu Lee and Won-Gi Paeng. This note is largely based on the work done with them in the World Class Program at Hanyang University, Seoul, Korea.


\begin{thebibliography}{50}

\bi{bsHLS} H.~Dong, T.~T.~S.~Kuo, H.~K.~Lee, R.~Machleidt and M.~Rho,
  ``Half-Skyrmions and the Equation of State for Compact-Star Matter,''
  Phys.\ Rev.\ C {\bf 87}, 054332 (2013);  W.~G.~Paeng, T.~T.~S.~Kuo, H.~K.~Lee and M.~Rho,
  ``Scale-Invariant Hidden Local Symmetry, Topology Change and Dense Baryonic Matter,''
  Phys.\ Rev.\ C {\bf 93}, no. 5, 055203 (2016);
  W.~G.~Paeng, T.~T.~S.~Kuo, H.~K.~Lee, Y.~L.~Ma and M.~Rho,
  ``Scale-invariant hidden local symmetry, topology change, and dense baryonic matter. II.,''
  Phys.\ Rev.\ D {\bf 96}, no. 1, 014031 (2017);    
 
  \bibitem{1.97S} P. Demorest {\it et al.}   ``Shapiro delay measurement of a two solar mass neutron star,"  {
  Nature\ {\bf 467},  1081 (2010).}
\bibitem{antoniadis13}  J. Antoniadis {\it et al.},
``A massive pulsar in a compact relativistic binary," {
  Science {\bf 340}, 6131 (2013)}

\bi{kaplan-nelson} D.B Kaplan and A.E. Nelson,   
``Kaon condensation in dense matter," 
  Nucl.\ Phys.\ A {\bf 479}, 273c. (1988).
  
\bi{BLR} G.E. Brown, C.H. Lee,  and M. Rho,  
  ``Recent developments on kaon condensation and its astrophysical implications, "{
  Phys.\ Rept.\  {\bf 462}, 1 (2008).}
  
   

\bi{pandharipande-thorsson} V.R. Pandharipande, C.J. Pethick,  and V. Thorsson, 
 `` Kaon energies in dense matter,"  {
  Phys.\ Rev.\ Lett.\  {\bf 75}, 4567 (1995).}

\bi{LR-hk} H.K. Lee and M. Rho,
``Hyperons and condensed kaons in compact stars," {
  arXiv:1301.0067 [nucl-th].}

 
\bi{callan-klebanov} C.G. Callan Jr. and I.R. Klebanov,
  ``Bound state approach to strangeness in the Skyrme model," {\it
  Nucl.\ Phys.\ B {\bf 262}, 365 (1985).}
  
\bi{klebanov-cargese-lecture} I.R. Klebanov,  ``Strangeness in the Skyrme model, "{\it Cargese Lecture on Hadron and Hadronic Matter, Cargese, France, 1989.}

\bi{Klebanov-crystal} I.R. Klebanov, 
 ``Nuclear matter in the Skyrme model,"  { Nucl.\ Phys.\ B {\bf 262}, 133 (1985)}.  
 
\bibitem{bedaque-steiner} P. Bedaque and A.W. Steiner, 
 ``Hypernuclei and the hyperon problem in neutron stars, "{
  Phys.\ Rev.\ C {\bf 92},  025803 (2015).}
  

\bi{BR:DD} G.E. Brown and M. Rho, 
 ``Double decimation and sliding vacua in the nuclear many body system," {
  Phys.\ Rept.\  {\bf 396}, 1 (2004).}
  
 \bi{CND-III} Y.L. Ma and M. Rho, {\it Effective Field Theories, Dense Matter and Compact Stars}\ (World Scientific, Singapore, 2018).
 
 \bi{gA} Y.~L.~Li, Y.~L.~Ma and M.~Rho,
  ``Nuclear axial currents from scale-chiral effective field theory,''
  arXiv:1710.02840 [nucl-th]. 
 \bi{FR96} B. Friman and M. Rho, 
  ``From chiral Lagrangians to Landau Fermi liquid theory of nuclear matter," {
  Nucl.\ Phys.\ A {\bf 606},  303 (1996).}
  
   \bibitem{LPR:PR15}  H.K. Lee, W.G. Paeng and M. Rho, 
 ``Scalar pseudo-Nambu-Goldstone boson in nuclei and dense nuclear matter," {
  Phys.\ Rev.\ D {\bf 92}, no. 12, 125033 (2015).}

 \bi{kojoetal}  G.~Baym, T.~Hatsuda, T.~Kojo, P.~D.~Powell, Y.~Song and T.~Takatsuka,
  ``From hadrons to quarks in neutron stars,''
  arXiv:1707.04966 [astro-ph.HE]; 
  T.~Kojo, P.~D.~Powell, Y.~Song and G.~Baym,
  ``Phenomenological QCD equations of state for neutron stars,''
  Nucl.\ Phys.\ A {\bf 956}, 821 (2016).
  
 \bi{Li-Ma-Rho} Y.~L.~Li, Y.~L.~Ma and M.~Rho,
  ``Chiral-scale effective theory including a dilatonic meson,''
  Phys.\ Rev.\ D {\bf 95},  114011 (2017).  
  
 \bi{Li-Ma}   Y.~L.~Li and Y.~L.~Ma,
  ``Derivation of Brown-Rho scaling from scale-chiral perturbation theory,''
  arXiv:1710.02839 [hep-ph].  
  
  

 \bi{half-skyrmions}  H.J. Lee,  B.Y. Park,  D.P. Min, M. Rho and V. Vento, 
  ``A unified approach to high density: Pion fluctuations in skyrmion matter, "{
 Nucl.\ Phys.\  A {\bf 723}, 427 (2003).}
 
\bi{song-brown-min-rho} C. Song,  G.E. Brown,  D.P. Min,  and M. Rho,  
  ``Fluctuations in 'BR scaled' chiral Lagrangians," {
  Phys.\ Rev.\ C {\bf 56}, 2244 (1997).}


 \bi{BTKR} G.E. Brown,  V. Thorsson, K. Kubodera and M. Rho, 
  ``A Novel mechanism for kaon condensation in neutron star matter," {
  Phys.\ Lett.\ B {\bf 291}, 355 (1992).}


\bi{PR-fermiliquid}  W.G. Paeng and M. Rho,  
  ``Kaon condensation in baryonic Fermi liquid at high density," {
  Phys.\ Rev.\ C {\bf 91}, 015801 (2005).}
  
 
\bi{holt-fermiliquid} J.W. Holt,  G.E. Brown,  J.D. Holt and  T.T.S. Kuo, 
  ``Nuclear matter with Brown-Rho-scaled Fermi liquid interactions," {
  Nucl.\ Phys.\ A {\bf 785},   322 (2007).}
  
  \bi{LRS} H.K. Lee,  M. Rho and S.J. Sin, 
``Renormalization group flow analysis of meson condensations in dense matter," {
  Phys.\ Lett.\ B {\bf 348}, 290 (1995).}

\bi{kachru} A.L. Fitzpatrick,  S. Kachru,  J. Kaplan and S. Raghu, 
  ``Non-Fermi liquid fixed point in a Wilsonian theory of quantum critical metals," {
  Phys.\ Rev.\ B {\bf 88}, 125116 (2013).}



   
  
\bi{MR-prl91} M.  Rho,   ``Exchange currents from chiral Lagrangians," {Phys,\ Rev.\ Lett.\ {\bf 66}, 1275 (1991).}



 \bi{shankar} R. Shankar,   ``Renormalization group approach to interacting fermions," 
  Rev.\ Mod.\ Phys.\  {\bf 66}, 129 (1994).
  
 \bi{confined halfskyrmions} S.~B.~Gudnason and M.~Nitta,
  ``Fractional Skyrmions and their molecules,''
  Phys.\ Rev.\ D {\bf 91},  085040 (2015);  P.~Zhang, K.~Kimm, L.~Zou and Y.~M.~Cho,
  ``Re-interpretation of Skyrme theory: New topological structures,''
  arXiv:1704.05975 [hep-th];  
 \bi{Vlowk}  S.K. Bogner,  T.T.T.S. Kuo and A. Schwenk, 
  ``Model independent low momentum nucleon interaction from phase shift equivalence," 
  Phys.\ Rept.\  {\bf 386}, 1 (2003).
  \bi{sigmaterm}  C. Alexandrou, 
``Nucleon structure from lattice QCD - recent achievements and perspectives," 
  EPJ Web Conf.\  {\bf 73}, 01013 (2014).
  
 \bi{gelmini-ritzi} G. Gelmini and B. Ritzi,  ``Chiral effective Lagrangian description of bulk nuclear matter,"
  Phys. Lett. B {\bf 357}, 431 (1995).
  
  
  \end{thebibliography}
\end{document}